\documentclass[prl,aps,nofootinbib,showkeys,showpacs,twocolumn,floatfix]{revtex4}   
\usepackage{epsfig}  
\usepackage{graphicx} 
\usepackage{amsmath}
\usepackage{amssymb}
\usepackage{amsfonts}
\usepackage{color}
\usepackage{todonotes}
\usepackage{braket}
\usepackage{color}  
\begin{document}  

\date{\today}
\title{
Ultrafast dynamics of finite Hubbard clusters -- a stochastic mean-field approach\\}

\author{Denis Lacroix}
\affiliation{Institut de Physique Nucl\'eaire, IN2P3-CNRS, Universit\'e Paris-Sud, F-91406 Orsay Cedex, France }
\author{S.~Hermanns, C.M.~Hinz, and M.~Bonitz}
\affiliation{Christian-Albrechts-Universit\"at zu Kiel, Institut f\"ur Theoretische Physik und Astrophysik, Leibnizstra{\ss}e 15, 24098 Kiel, Germany}

\begin{abstract}
Finite lattice models are a prototype for strongly correlated quantum systems and capture essential 
properties of condensed matter systems. With the dramatic progress in ultracold atoms in optical lattices, finite fermionic Hubbard systems have become 
directly accessible in experiments, including their ultrafast dynamics far from equilibrium. Here, we present a theoretical
approach that is able to treat these dynamics in any dimension and fully includes inhomogeneity effects.
The method consists in stochastic sampling of mean-field trajectories and is found to be more accurate and efficient than 
current nonequilibrium Green functions approaches. This is demonstrated for Hubbard clusters with up to 512 particles in one, two 
and three dimensions.
%in sampling the initial phase-space of one-body degrees of freedom followed by a set of mean-field evolutions is applied 
%to study the non-equilibrium evolution of fermions in the Hubbard one- and three-dimensional model. The transport theory is shown 
%to properly treat the propagation of initial quantum fluctuations and their effect on one-body degrees of freedom 
%for not too strong interactions. It turns out to be competitive compared to 
%other methods like Non-Equilibrium Green-Functions without facing the explosion of degrees of freedom as the system size increases.
\end{abstract}
\pacs{71.10Fd,71.27.+a,74.40.-n}
\keywords{Lattice models, mean-field, stochastic methods}
\maketitle

Experimental progress in the formation and manipulation of quantum optical lattices from one to three dimensions with the number of sites ranging from very few to 
thousands provide a perfect laboratory \cite{Blo08,Hac12,Sta12,Sto10} for the study of nonequilibrium properties of mesoscopic bosonic and fermionic systems.  Exciting recent observations include the formation and expansion of fermionic pairs (doublons) in a lattice, e.g. \cite{kajala_prl11, kessler_njp13}, the many-particle behavior following an interaction quench \cite{bernier_prl14} or the transport behavior following a quench of the confinement potential \cite{schneider_np12}.
This paves the way towards a fundamental understanding of strongly interacting quantum systems, which is of prime importance for many areas of physics and chemistry, including ultracold atomic and molecular gases in traps and optical lattices,  transport and coherence properties of macromolecules, superconductivity and magnetic properties of condensed matter systems 
on the nanoscale, thermodynamics of dense fermionic matter in compact stars and so on.

The description of finite  correlated quantum lattice systems out of equilibrium is very challenging.
Exact solutions using direct configuration interaction (CI) methods are possible only for very small
Hubbard clusters~\cite{Akb08}, and 
%However, even with most recent computers, only rather small 
 the use of Quantum Monte-Carlo methods \cite{Fou01} only slightly increases the accessible system size. These approaches are hampered by the exponential increase of the computational effort with the system size. 
%Furthermore, these approaches are not flexible enough to approach most of the current experimental situations. 
%
For this reason, the interest in approximate nonequilibrium theories that are both reliable and efficient
%able to tackle the problem of quantum lattice out-of equilibrium 
has recently increased substantially. The simplest approach to treat particle-particle interactions is via an effective average field, i.e., via mean-field theory (e.g. time-dependent Hartree-Fock (TDHF)). However, with increasing quantum entanglement, this approximation quickly fails. Moreover, as we will show below [cf. Fig.~\ref{fig1:lattice}], 
%in nonequilibrium,
 TDHF already fails for very small coupling strength (Hubbard-$U$). 

To describe the systems and phenomena mentioned above, one has to resort to methods beyond mean-field, i.e., include correlation effects. In recent years, there has been remarkable progress in this direction, in particular, in the application to strongly correlated lattice systems. Among the successful approaches,
%that extends the dynamical mean-field theory (DMFT {\em Micha: DMFT is well defined term for DFT plus some correlations. We should use TDHF instead}) have been proposed.  
we mention the time-dependent density matrix renormalization group method (TDDMRG), e.g. \cite{bernier_prl14}, time-dependent density functional theory (TDDFT, e.g. \cite{Fuk13}), Nonequilibrium Green Functions (NEGF), e.g. \cite{Bon98, dahlen07, puigvonfriesen09}, or time-dependent density matrix (TDDM) methods \cite{Bon98, Akb12}. All these methods have various limitations, e.g. with respect to the correlation strengths (TDDFT, TDDM, NEGF) or the system dimensionality (TDDMRG). Recent benchmarks for the 1D Fermi--Hubbard model indicated fundamental problems such as unstable behavior of TDDM \cite{Akb08} or unphysical damping in NEGF simulations \cite{puigvonfriesen09} that could recently be overcome in part by applying the generalized Kadanoff--Baym ansatz (GKBA) \cite{Lip86, Bon13, Her13, Her14}. At the same time, going beyond the mean-field level with these approaches is rather involved and very expensive in terms of computational resources.

%where two-body Degrees of Freedom (DOF) are followed in time.  
%Recently, these two approaches have been benchmarked 
%on the one-dimensional Hubbard model. 
%Using the generalized Kadanoff-Baym ansatz (GKBA)  
%or several truncation scheme of the so-called BBGKY hierarchy \cite{Bon98, Akb12} gives significant improvements beyond the simple
%dynamical mean-field picture. These theories are rather involved and even if the exponential explosion with the number of sites $N$ is significantly 
%reduced, 
%the number of DOF to follow in time prevents from using them on a systematic basis. Conjointly, deviation between 
%state of the art NEGF in the GKBA approximation and exact evolution are systematically showing that some important correlations are still missing even in the weak coupling regime. 

%The possibility to describe fast evolution of strongly interacting system while keeping the simplicity of mean-field is challenging the field. One attractive approach is the Time-Dependent Density Functional Theory (TDDFT) but for lattice model it is still at the exploratory stage (see Ref. \cite{Fuk13}). 

%In this Letter, w
We propose an alternative approach to correlated fermionic lattice systems where the simplicity of mean-field theory is combined with efficient stochastic methods that allow to incorporate correlation effects.
This {\em Stochastic Mean Field} approach (SMF) has been recently developed and applied with success in nuclear physics \cite{Ayi08,Lac12,Lac13,Lac14}. We present tests against exact results for small lattice systems and demonstrate that 
SMF is accurate for weak to moderate coupling during the initial phase of relaxation. Moreover, applying it to large 1D, and to 3D systems---where no CI data are available---we demonstrate its impressive capabilities for extended systems.

%{\em Micha: I propose to note the nuclear theory work here already but without the details of the initial state sampling.}
%a proper sampling of the initial quantal phase-space. 
%
\begin{figure}[!ht]
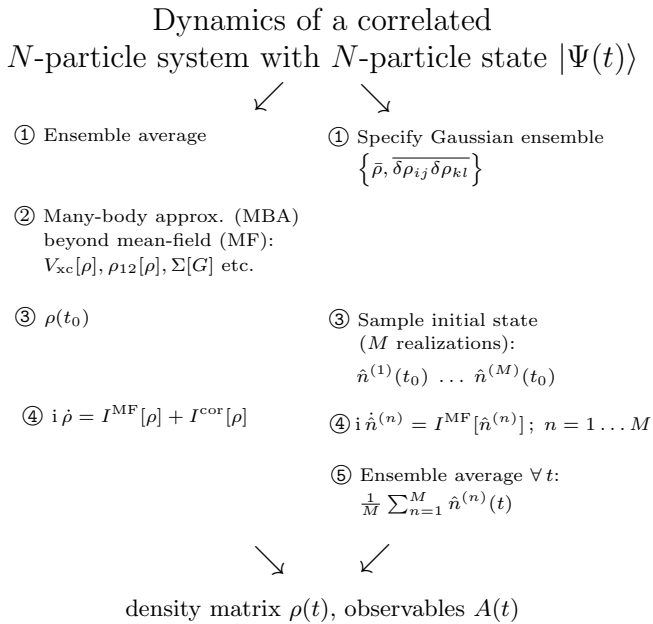

\begin{center}
\large
Dynamics of a correlated\\
$N$-particle system with $N$-particle state $\ket{\Psi(t)}$
\end{center}
\vspace*{-0.3cm}
\begin{minipage}{0.49\columnwidth}
\scriptsize
% \vspace*{-0.4cm}
\hspace*{3cm}{\large$\swarrow$}\\[0.2cm]
\hspace*{-1.33cm} {\footnotesize\textcircled{\texttt{\raisebox{-0.05em}{\scriptsize 1}}}}\raisebox{-0.05em}{\hspace*{0.1cm}Ensemble average}\\[0.8cm]
{\footnotesize \textcircled{{\raisebox{-0.05em}{\scriptsize 2}}}}\raisebox{-0.05em}{ Many-body approx. (MBA)}\\[0.05cm]
\hspace*{0.06cm}beyond mean-field (MF):\\[0.05cm]
\hspace*{-0.2cm}$V_\textnormal{xc}[\rho], \rho_{12}[\rho], \Sigma[G]$ etc.\\[0.4cm]
\hspace*{-2.78cm}{\footnotesize\textcircled{\texttt{\raisebox{-0.05em}{\scriptsize 3}}}} \raisebox{-0.05em}{\hspace*{0.01cm}$\rho(t_0)$}\\[1.5cm]
\vspace*{-0.5cm}
\hspace*{-0.58cm}{\footnotesize\textcircled{\texttt{\raisebox{-0.05em}{\scriptsize 4}}}} \raisebox{-0.05em}{\hspace*{0.01cm}$\textnormal{i}\,\dot{\rho}=I^{\textnormal{MF}}[\rho]+I^{\textnormal{cor}}[\rho]$}\\[2.05cm]
\vspace*{-0.5cm}
\hspace*{3cm}{\large$\searrow$}\\[-0.9cm]
\end{minipage}
\hfill
\begin{minipage}{0.49\columnwidth}
\scriptsize
\hspace*{-3.0cm}{\large$\searrow$}\\[0.2cm]
\hspace*{-0.60cm}{\footnotesize\textcircled{\normalsize\texttt{\raisebox{-0.05em}{\scriptsize 1}}}} \raisebox{-0.05em}{Specify Gaussian ensemble}\\[0.05cm]
\hspace*{-1.8cm}$\left\{\bar{\rho},\overline{\delta \rho_{ij}\delta \rho_{kl}}\right\}$\\[1.65cm]
\hspace*{-1.47cm}{\footnotesize\textcircled{\normalsize\texttt{\raisebox{-0.05em}{\scriptsize 3}}}} \raisebox{-0.05em}{Sample initial state}\\[0.05cm]
\hspace*{-1.36cm}($M$ realizations):\\[0.1cm]
\hspace*{-0.85cm}$\hat{n}^{(1)}(t_0)\;\ldots\;\hat{n}^{(M)}(t_0)$\\[0.80cm]
\vspace*{-0.5cm}
{\footnotesize\textcircled{\normalsize\texttt{\raisebox{-0.05em}{\scriptsize 4}}}}\raisebox{-0.05em}{\hspace*{0.05cm}$\textnormal{i}\,\dot{\hat{n}}^{(n)}=I^{\textnormal{MF}}[\hat{n}^{(n)}]\,;\; n=1\ldots M$}\\[0.4cm]
\hspace*{-1.15cm}{\footnotesize\textcircled{\normalsize\texttt{\raisebox{-0.05em}{\scriptsize 5}}}} \raisebox{-0.05em}{Ensemble average $\forall\,t$:}\\[0.05cm]
\hspace*{-1.4cm}$\frac{1}{M}\sum_{n=1}^M \hat{n}^{(n)}(t)$\\[0.85cm]
\vspace*{-0.5cm}
 \hspace*{-3cm}{\large$\swarrow$}
\end{minipage}
 \begin{center}
 density matrix $\rho(t)$, observables $A(t)$
\end{center}
        \caption{Comparison of the strategy of typical many-body approaches (left) to the present stochastic mean-field method (right). Many-body approximations are specified, e.g., by the exchange-correlation potential $V_{\rm xc}$ (DFT), by the functional dependence of the two-body density matrix $\rho_{12}$ on $\rho$ or the form of the selfenergy $\Sigma$ (NEGF theory). In contrast, for the SMF, only TDHF trajectories are used and correlations are mimicked by a properly chosen ensemble.}
         \label{fig1:scheme}
\end{figure}
% \begin{figure}[!ht]
% \todo{Rescale figure 1}
%         \centering\includegraphics[height=8cm,width=0.95\linewidth]{scheme2.png}  
% %        \centering\includegraphics[width=1.1\linewidth]{fig1lattice.eps}  
%         \caption{Comparison of the strategy of typical many-body approaches (left) to the present stochastic mean-field method (right). Many-body approximations are specified, e.g., by the exchange-correlation potential $V_{\rm xc}$ (DFT), by the functional dependence of the 2-body density matrix $\rho_{12}$ on $\rho$ or the form of the selfenergy $\Sigma$ (NEGF theory). In contrast, for the SMF, only TDHF trajectories are used and correlations are mimicked by a properly chosen ensemble.}
%          \label{fig1:scheme} 
% \end{figure}

\noindent
{\bf Beyond mean-field transport theories.} 
Most approaches going beyond the independent particle picture start from a generalized one-body equation of motion (EOM) where 
the effect of correlations, associated to the correlation matrix $C_{12}$ is accounted for. For an ensemble of particles interacting through a (anti-)symmetrized two-body interaction $\tilde v_{12}$, the 
exact evolution of the one-body density matrix reads ($[\cdot , \cdot\cdot]_-$ denotes the commutator and $\hbar\equiv1$):
\begin{eqnarray}
\mathrm{i} \partial_t \rho & = & \left[ h(\rho), \rho\right]_- +\frac{1}{2} {\rm Tr}_2 \left[\tilde v_{12}, C_{12}\right]_-  \label{eq:rhoexact} \\
&=:&  I^{\rm MF} \left[\rho \right] + I^{\rm cor}  \left[ C_{12} \right]\,, 
\end{eqnarray} 
where $h(\rho)= T+V+ {\rm Tr_2} (\tilde v_{12} \rho_2) =: T+V+ U_{\rm HF}(\rho) $ is the mean-field Hamiltonian containing kinetic ($T$) and potential ($V$) energy and the potential $U_{\rm HF}$ that is induced by all particles. At the mean-field level, correlations are neglected leading to the TDHF theory.  $C_{12}$, which is defined through the binary density operator via $\rho_{12} = \rho_1 \rho_2 (1\pm P_{12})+C_{12}$ [$P_{12}$ is the pair permutation operator], is---in general---unknown.
Theories that go beyond mean-field typically introduce approximations 
 for the correlation matrix, $C_{12}=C_{12}[\rho]$ (e.g. TDDM), or the exchange-correlation potential $V_{xc}$ (DFT). Similarly, within NEGF, corrections to TDHF are incorporated through
 the correlation selfenergy $\Sigma[G]$. The standard strategy to improve mean-field is presented schematically in Fig. \ref{fig1:scheme} (left part). 
%The explicit treatment of two-body degrees of freedom 
%is generally restricted to rather simple systems due to the size of the object to be propagated in time.  In some cases, approximations 
%of the correlation kernel can lead to a close EOM for the one-body density, i.e. $I^{\rm cor}  \left[ C_{12} \right] $ becomes a functional of $\rho$ itself.
Inclusion of correlation effects via the collision integral $ I^{\rm cor}$---even in a simplified version---in general leads to a much harder to solve problem compared to the original mean-field approximation. This becomes even more severe in nonequilibrium. Therefore, most standard approaches lead to a dramatic increase of the computational complexity of the problem and/or put significant 
restrictions on the system size or coupling strength that can be treated. 
%reducing significantly their versatility.  

%\noindent
{\bf Stochastic Mean Field Theory} presents a fundamentally different strategy that aims at incorporating correlations (at least partially) while retaining the simplicity of a mean-field description. 
We consider a stochastic scheme where an ensemble of single-particle density matrices $n^{(n)}(t)$ is used, where $(n)$ labels a given realization (trajectory). In the following, we denote by $\overline{ n}(t)=(1/M)\sum_M n^{(n)}(t)$
the average over the $M$ trajectories and by $\delta n^{(n)}(t) = n^{(n)}(t) - \overline{ n}(t)$ the individual fluctuations around this mean. Each density is assumed to evolve according to 
its own mean-field dynamic:
 \begin{eqnarray}
\label{eq:smf0}
 \mathrm{i} \partial_t n^{(n)} & = & I^{\rm MF} \left[n^{(n)}\right]=\left[ h(n^{(n)}), n^{(n)}\right]_-\,,\\
n^{(n)}(t_0)& = &n^{(n)}_0\,.\nonumber
\end{eqnarray}
A straightforward derivation shows that the evolution of the average density is given by
\begin{eqnarray}
\mathrm{i} \partial_t \overline{n} = I^{\rm MF} \left[\overline{n} \right] + \overline{ \left[ \delta U_{\rm HF}(n^{(n)}) , \delta n^{(n)} \right]_-}
\end{eqnarray} 
where $\delta U_{\rm HF}(n^{(n)})$ denotes fluctuations of the induced potential introduced by the density fluctuations. 
Comparing this average evolution with Eq. (\ref{eq:rhoexact}), we see that evolving a statistical ensemble of densities can simulate the effect of correlations \cite{comment} provided that $\overline{n} (t) = \rho(t)$ and  
\begin{eqnarray}
\lim_{M \rightarrow \infty} \frac{1}{M}\sum_{n=1}^M \overline{[ \delta U_{\rm HF}(n^{(n)}) , \delta n^{(n)} ]_-}  &=&  I^{\rm cor}  \left[ C_{12} \right].  \label{eq:const}
\end{eqnarray}
 
This correspondence is exact if condition (\ref{eq:const}) is fulfilled for all times, however, the correlation function 
$\overline{[ \delta U_{\rm HF}(n^{(n)}) , \delta n^{(n)} ]_-}$ is of similar complexity as the pair correlation matrix $C_{12}$.
To overcome this problem, the present stochastic mean-field approach proposes to map the time-dependent correlations onto fluctuations at the initial time $t_0$.
%In practice, initial fluctuations are chosen in such a way that, at initial time $t_0$, the mean values and fluctuations of one-body observables 
Furthermore, the fluctuation spectrum is chosen such that it matches the quantum expectation value and fluctuations of the one-body density matrix (OBDM) of the initial state of the system (Gaussian approximation, details will be explained below for the Hubbard model). Then, each randomly chosen initial density is propagated in its own mean-field through 
Eq.  (\ref{eq:smf0}) as is illustrated in the right part of Fig.~\ref{fig1:scheme}. In this manner, we anticipate to achieve---at least for weak entanglement---correct correlated dynamics. 
%
%\begin{figure}[!ht]   
%\vspace*{-3cm}

\begin{figure}[!b]   
        \centering\includegraphics[width=.9\linewidth]{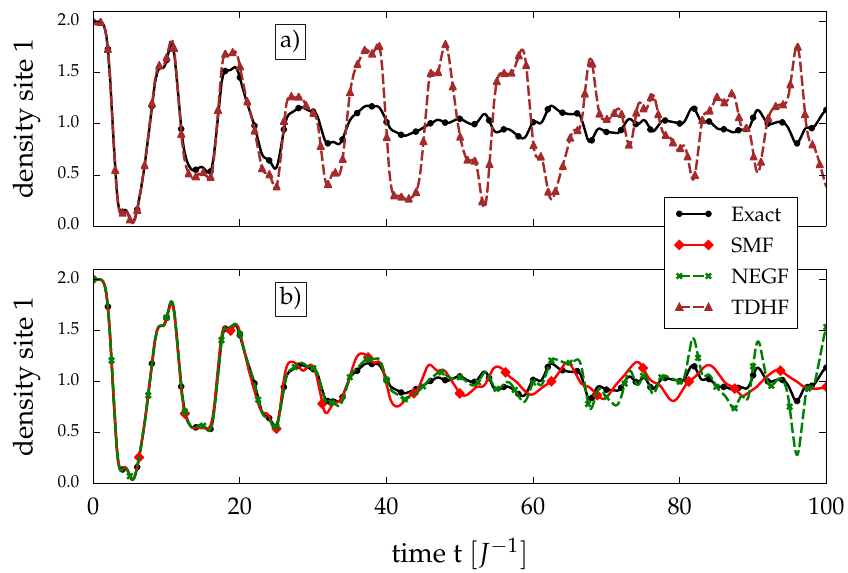}  
%        \centering\includegraphics[width=1.\linewidth]{fig1lattice.pdf}  
%        \centering\includegraphics[width=1.1\linewidth]{fig1lattice.eps}  
%\vspace*{-2cm}
        \caption{(Color online) Time evolution of the leftmost site occupation $n_1(t)$ for $N_s=N=8$ and $U/J = 0.1$. 
        a) CI (solid line) vs. TDHF (dashes, triangles) and b) CI vs. NEGF (green, dashes) and SMF (red, diamonds).}
         \label{fig1:lattice} 
\end{figure}
A proof that such an approach is able to describe correlated quantum lattice systems is challenging because mean-field per se is a poor approximation, even in the weak coupling limit (see top panel of Fig. \ref{fig1:lattice} and discussion below). 
It is the aim of this Letter to provide numerical evidence and to map out the parameter range where this approach works.
In fact, we demonstrate that our SMF approach is surprisingly accurate at weak to moderate coupling. Most importantly, since only standard mean-field evolution is needed, SMF does not face the catastrophic explosion of computational effort with system size as standard many-body theories.  
%DOF to follow in time occurring in other approaches like NEGF or TDDM.  

%\noindent
{\bf Application to fermionic Hubbard clusters.}
As an illustration of the capabilities of the SMF approach, a $d$-dimensional Hubbard nano-cluster consisting of $N$ doubly degenerate sites is considered. The Hamiltonian in the canonical Hubbard basis reads:
\begin{eqnarray}
H = -J \sum_{i,j}^N\sum_\sigma\delta_{<i,j>} c^\dagger_{i\sigma}  c_{j \sigma} 
+ U \sum_i^N c^\dagger_{i\uparrow} c^\dagger_{i \downarrow}\,.  
c_{i \downarrow} c_{i \uparrow}\,,
\end{eqnarray}
where $\delta_{<i,j>}=1$ for nearest-neighbor sites and equals zero otherwise. Note that the dimensionality---and any boundary conditions---enter the Hamiltonian only via specifying which sites are nearest neighbors. 
%In the following, the interaction strength $U$ and time will be given in units of $J$
%and $J^{-1}$, respectively. 
This idealized Hamiltonian describes the interplay between strong localization and fast 
tunneling and was placed on the front of the stage by recent progress in cold atoms in optical lattices \cite{Jak98,Gre02}.  Mean-field dynamics provide a set of coupled equations for the OBDM $\rho_{i\sigma j\tau} = \langle c^\dagger_{j\tau}  
c_{i \sigma} \rangle $, where $\sigma,\tau$ denote the spin orientations. For a spin symmetric system, the spin variables can be omitted, and we simply use the notation $\rho_{ij}=\rho_{i\uparrow j\uparrow}=\rho_{i\downarrow j\downarrow}$. 

%The observables $\{ c^\dagger_{j,\sigma_j}  
%c_{i \sigma_i} \}$ could be regarded as the set of DOF evolving according to coupled equations of the form (\ref{eq:allobs}).  
Within the SMF scheme, the initial OBDM is replaced by 
an ensemble of initial realizations, i.e. $\rho(t_0) \rightarrow \{ n^{(n)}(t_0) \}$, each of which evolves according to Eq.~(\ref{eq:smf0})
with the mean-field Hamiltonian 
%
%\begin{eqnarray}
$h\left(n^{(n)}\right)_{ij}=-J \delta_{<i,j>} +U\delta_{ij}n^{(n)}_{ii}.$ 
%\end{eqnarray}
%
The (random) initial configurations $\{ n^{(n)}(t_0) \}$ in the Hubbard basis are determined by demanding that the matrix elements $\tilde{n}^{(n)}_{ij}(t_0)$ of $n^{(n)}(t_0)$ satisfy 
\begin{eqnarray}
\overline{\tilde{n}^{(n)}_{ij}(t_0)} &=& \delta_{ij}\tilde{n}_i(t_0) , \\
\overline{\delta \tilde{n}^{(n)}_{ij} (t_0)\delta \tilde{n}^{(n)}_{kl}(t_0)} &=&  \frac{1}{2} \delta_{jk} \delta_{il} \tilde{n}_j(t_0) \left(1 -\tilde{n}_i(t_0)\right)
\end{eqnarray}
in the natural orbital basis of $\rho(t_0)$ with $\delta \tilde{n}^{(n)}_{ij} = \tilde{n}^{(n)}_{ij} - \overline{\tilde{n}^{(n)}_{ij}}$, where $\tilde{n}_i(t_0)$ denote the eigenvalues of $\rho(t_0)$. 
%Here, $n_i$ denotes the occupation number of the initial quantum state. 
Here, only the first two moments of the density are constrained corresponding to a Gaussian ensemble (constraining higher moments as well may provide additional flexibility). At any time $t$, the expectation values of the site occupations can be computed simply using $n_i(t)=\overline{n^{(n)}_{ii}(t)}$, whereas fluctuations of the occupations become accessible by $\sigma^2_{i}=\overline{\delta {n^{(n)}_{ii}} \delta { n^{(n)}_{ii}}} $.
%
%between two observables follow from $\overline{ \delta Q^{(n)}  \delta P^{(n)} } = \sum_{ij,kl} Q_{ij} P_{kl} \overline{ \delta n^{(n)}_{ji}\delta n^{(n)}_{lk}}$.

A first illustration is shown in Fig. \ref{fig1:lattice} for a half-filled 8-site 1D-chain without periodic boundary conditions at weak two-body interaction where we compare the exact solution and approximations. The initial state is chosen such that all 8 electrons reside on the left-most four sites. 
We see that even in the weak coupling regime, the mean-field result (TDHF, Fig.~\ref{fig1:lattice}.a) deviates very fast from the exact case and is, in particular, unable to describe the damping of the site occupation. 
In Fig.~\ref{fig1:lattice}.b, we also show the NEGF-GKBA result \cite{Her14} which performs much better (green, dashes). 
Finally, the SMF result obtained by using $10^4$ fluctuating initial conditions is shown (red, diamonds). Obviously, SMF 
provides a dramatic improvement over TDHF, properly describes the short time damping but also reasonably well describes  the long time dynamics until about $t=75 [J^{-1}]$. 
%
%\begin{figure}[!ht]   
\begin{figure}[h]   %        \centering\includegraphics[width=1.1\linewidth]{fig2lattice.eps}  
%        \centering\includegraphics[width=1.\linewidth]{fig2lattice.pdf}  
        \centering\includegraphics[width=.9\linewidth]{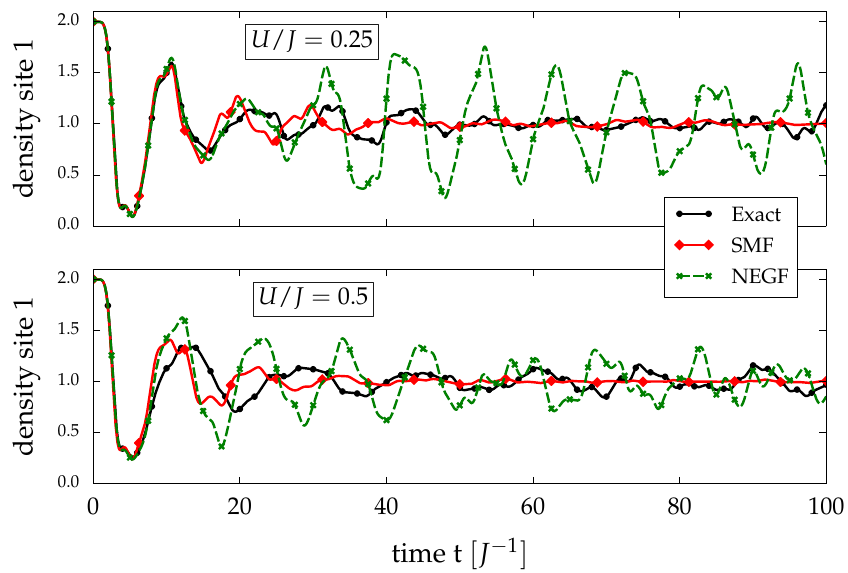}  
%        \vspace*{-2cm}
        \caption{(Color online)  Time evolution of the leftmost site occupation $n_1(t)$ for $N_s=N=8$, for $U/J=0.25$ (top) and $U/J=0.5$ (bottom). The exact solution (black solid line), NEGF (green, dashes) and SMF (red, diamonds) are shown.}
         \label{fig2:lattice} 
\end{figure}
In particular, the revival observed at long times in the NEGF result ($t\sim 80, 90, 95  [J^{-1}]$), are absent, in agreement with the exact case. For all particle numbers where CI data are available, in the weak coupling regime, we observed a similar accuracy of SMF, and, in most cases, the results are also more accurate than the NEGF ones. When the interaction $U/J$ increases, the overall behavior is still correctly reproduced, in contrast to TDHF. At the same time,  quantitative agreement with CI is restricted to shorter times, cf. Fig. \ref{fig2:lattice}. Interestingly, the validity range seems to be bounded by the correlation time of the system \cite{Her14}, $t \lesssim \tau_{cor} \sim 1/U$. 

\begin{figure}[!ht]   
%       \centering\includegraphics[width=0.7\linewidth,angle =-90]{fig3lattice.eps}  
%        \centering\includegraphics[width=0.9\linewidth,angle = 0]{fig3lattice.pdf}  
        \centering\includegraphics[width=0.8\linewidth]{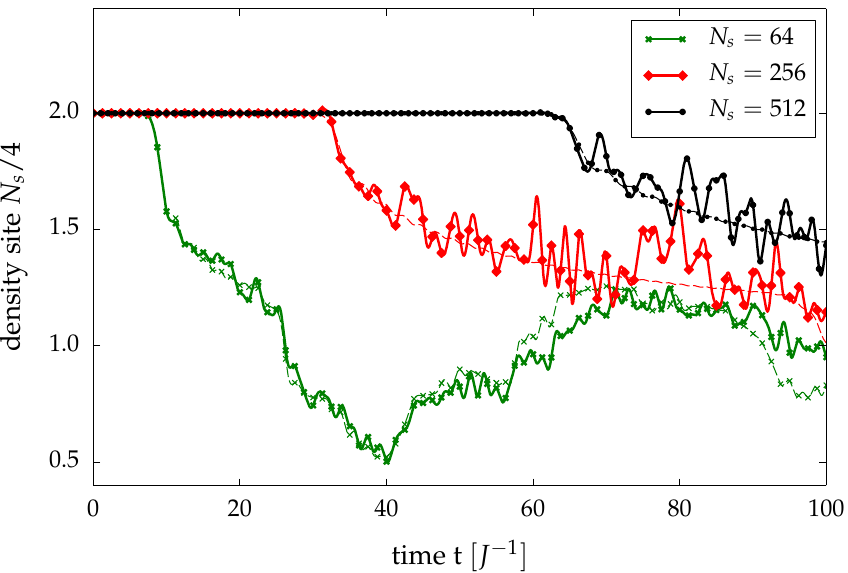}  
        \caption{(Color online)  SMF result for large one-dimensional Hubbard systems with   $U/J=0.1$ and half filling for $N_s=64$ (green), $N_s=256$ (red) and $N_s=512$ (black). The dynamics of the site occupation $n_{N_s/4}(t)$ are shown and compared to Hartree-Fock results (thin dashed lines).}
         \label{fig3:lattice} 
\end{figure}

%The failure of mean-field theory and the success of SMF clearly points out that the main source of deviation of mean-field from the exact case in the weak coupling regime 
%stems from its inability to propagate initial fluctuations in relevant DOF space.  Such fluctuation evolution could eventually be propagated through the development 
%of time-dependent RPA methods but this would be numerical costly. 
%We thus conclude that the present SMF approach is able to reliably describe Hubbard clusters far from equilibrium up to $U/J \sim 0.25$ over (at least) the first three oscillation cycles.
%while keeping the simplicity of mean-field, turns out to properly treat 
%the evolution and possible amplification of initial quantum fluctuations. 
%
Since only a mean-field like evolution is required, the numerical effort for SMF essentially scales as for TDHF and, therefore, SMF can be applied to cases where other methods would either require supercomputing facilities or cannot be applied at all. We illustrate this point in Fig. \ref{fig3:lattice} where we apply SMF to long Hubbard chains of $N_s=64$, $256$ and $512$ sites, respectively ($U/J=0.1$) and no exact solutions are available. 
Starting from the same initial state as above (all $N$ particles occupy the leftmost sites $1 \dots N_s/2$) %we observe that 
the occupation of the left sites first remains constant and then displays a rather complex evolution (Fig. \ref{fig3:lattice} displays the initially occupied site $N_s/4$). The increase of the time before site depletion is due to the Pauli principle that prevents hopping until surrounding sites become at least partially empty. This time increases almost linearly with $N_s$ indicating that, here, correlation effects are of minor importance.
Also, the decrease of the population $n_{N_s/4}$ is strongly reduced with increasing $N_s$ since---with the delay of depletion---particles initially located to the right have already undergone reflections at the right boundary and affect the originally occupied sites.
%{\it Denis: here we can finish depending on the result. Actually, I would be in favor of pushing the SMF method to some large $N$ limit. Since you [Sebastian] can access GPU, could 
%you please try to increase $N$ as much as you can?} 

\begin{figure}[!ht]   
        \includegraphics[width=0.9\linewidth,angle = 0]{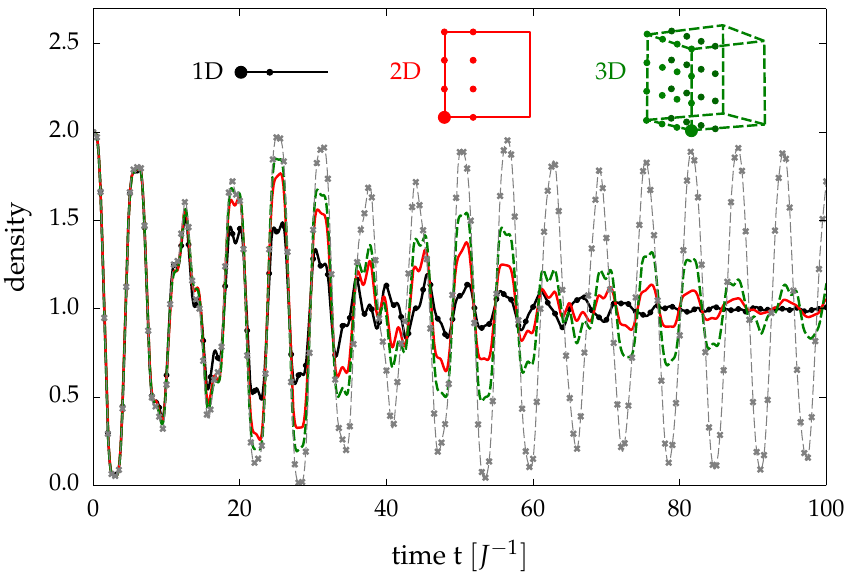} 
        \caption{(Color online)  SMF result for a three-dimensional Hubbard cluster of size $4\times 4\times 4$, compared to a 2D ($4\times 4$) and 1D ($N_s=4$) cluster (see inset for the initial state). The dynamics of the leftmost site (bold dot) occupation are shown and compared to the TDHF result (3D case, dashed line).
}
         \label{fig4:lattice} 
\end{figure}

As a final illustration, we demonstrate that the SMF method can be directly applied to clusters of higher dimensions.
In Fig.~\ref{fig4:lattice}, we show, as an example, a cubic arrangement of $64$ sites for half filling and weak coupling $U/J=0.1$. As the initial configuration, we place all particles to the left half. Comparison of SMF to TDHF shows, as before, that mean-field is not adequate and strongly underestimates the damping, although the dominant frequency is correctly captured. We now try to understand the effect of the dimensionality. To this end, we compare to a 2D cluster of $16$ sites (similar to a cut through the original cube) as well as to a linear chain of size $N_s=4$ (see inset of Fig.~\ref{fig4:lattice}). In all cases, the site occupations oscillate with almost the same main frequency. The most striking effect of dimensionality is that the damping of the oscillations grows when going from 3D to 1D. 

In summary, we studied the ultrafast dynamics of finite spatially inhomogeneous fermionic Hubbard clusters with up to $512$ sites in one, two and three dimensions driven far from equilibrium.
We applied a stochastic mean-field approach where first a phase-space sampling of collective degrees of freedom is performed, followed by a set of simple time-dependent mean-field evolutions. We demonstrated that SMF significantly improves the mean-field dynamics---which already fail at weak coupling---by incorporating the effect of initial fluctuations of one-body degrees of freedom.  Despite the fact that only a mean-field evolution is required, SMF can treat correlations and works very well, especially in the weak coupling regime. 
The main advantage is the very efficient inclusion of correlation effects via Monte Carlo sampling which overcomes the unfavorable exponential scaling of wave function based approaches and allows to access large systems of arbitrary dimensionality. Here, the approach is implemented in its simplest form, i.e. starting from 
an uncorrelated state, where correlations are incorporated via Gaussian fluctuations of the initial site occupations.
This, naturally, limits the method to the initial time stage, $t\lesssim \tau_{cor}\sim 1/U$, yet this regime is particularly interesting and difficult to treat since here correlations are being built up dynamically and Markovian approximations fail.
Extensions to longer times or/and stronger couplings seem to be straightforward, e.g. by relaxing the spin symmetry (unrestricted mean-field), by inclusion of pairing correlations \cite{Lac13} or via time-dependent fluctuations.
Finally, it is straightforward to extend the method to two-time fluctuations paving the way towards a stochastic approach to Nonequilibrium Green functions.

%We note that for couplings $U/J > 1$, the present version of SMF fails qualitatively. For this regime, further work is needed since first attempts to improve  have been hampered by the occurence of instabilities.

% extension to 
% NEGF is straightforward as well as the treatment of initial thermally equilibrated and/or correlated systems by proper initial fluctuation mapping.    
% An extension to larger couplings will require to include time-dependent fluctuations.
%


\begin{thebibliography}{99}
\bibitem{comment} The close relation of fluctuations to correlations has frequently been used to derive approximations for the correlation functions, e.g. in quantum electrodynamics \cite{schwinger51} or plasma physics \cite{klimontovich_book}. However, there it has not been used as a direct numerical tool.
\end{thebibliography}
\end{document}

\maketitle

%\begin{abstract}
%\end{abstract}

%\section{}

In the main text, we discuss two approaches to the time evolution of the one-body density matrix (see also Fig. 1 of the main text): The first is 
for the density matrix $\rho$, Eq.~(1). Here, $\rho$ is associated to the reduced one-particle 
density operator that is derived from the full $N$-particle density operator of the system [1], 
${\hat \rho} = N\mbox{Tr}_{2\dots N} {\hat \rho_{1\dots N}}$. Equation (1) is standard text book material (See Eq.~(3.20) of [1]).
The second approach is to express the correlation terms via fluctuations. Here, we outline the main steps 
leading to Eq. (4) of the main text.

We start with the one-particle density matrix operator ${\hat n}_{ij}\equiv {\hat a}_i^\dagger {\hat a}_j$ defined in 
terms of standard bosonic or fermionic creation and annihilation operators with respect to an arbitrary single-particle basis. Note that no ensemble averaging is applied yet.

The dynamics of ${\hat n}_{ij}$ is given by the Heisenberg equation
\begin{eqnarray}
 \mathrm{i}\partial_t\, {\hat n}_{ij}(t) = -\hat{U}^\dag(t,t_0)\left[{\hat H}(t),{\hat n}_{ij}\right]_-\hat{U}(t,t_0), \qquad {\hat n}_{ij}:={\hat n}_{ij}(t_0),
\label{eq:eom-ni}
\end{eqnarray}
 where $\hat{U}$ is the time evolution operator and ${\hat H}(t)$ the system Hamiltonian written in second quantization as
\begin{eqnarray}
 {\hat H} &=& {\hat T} + {\hat V} + {\hat W}\,,
\label{eq:h_2quant_ij}\\
 {\hat T} + {\hat V} &=& \sum_{ij} {\hat a}_i^\dag \left(t_{ij} + v_{ij}(t) \right) {\hat a}_j\,, 
%=  \sum_{i,j = 1}^\infty h_{ij}\,{\hat a}_i^\dag  {\hat a}_j, 
\label{eq:tu_2quant_ij}\\
  {\hat W} &=& \frac{1}{2}\sum_{ijkl} {\hat a}_i^\dag {\hat a}_j^\dag \, w_{ijkl} \, {\hat a}_l {\hat a}_k. 
\label{eq:w_2quant_ij}
\end{eqnarray}
Using these expressions and the (anti-)commutation relations of the operators ${\hat a}_i^\dagger$ and ${\hat a}_j$, the commutator in (\ref{eq:eom-ni}) is straightforwardly evaluated, with the result
\begin{eqnarray}
 \mathrm{i}\partial_t {\bf {\hat n}}(t) = \left[{\bf {\hat n}}(t),\left\{ {\bf t}^* + {\bf v}_{\rm H}^*(t) + {\bf {\hat U}}^{\pm}_{\rm H}(t) \right\} \right]_-\,.
%= \left[{\bf {\hat n}}(t), {\bf {\hat h}}^{\pm}_{\rm H}(t) \right]
%- \left\{ {\bf h}_{\rm H}^*(t) + {\bf {\hat U}}^{\rm ind}_{\rm H}\right\}{\bf n}(t)
\label{eq:eom-nij_final}
\end{eqnarray}
Here, we introduced bold notation for matrices of one-body operators, ${\bf {\hat n}}\equiv {\hat n}_{ij}$, and the 
subscript ``H'' designates a Heisenberg operator,\\ i.e. $\hat{A}_{\rm H} \equiv \hat{U}^\dagger(t,t_0) \hat{A}\, \hat{U}(t,t_0)$. Furthermore, 
the (anti-symmetrized) induced potential appearing in (\ref{eq:eom-nij_final}) is given by
\begin{eqnarray}
{\hat U}_{kj}^{\pm} &=& \sum_{ln} \frac{w_{jnkl} \pm w_{jnlk} }{2}\left\{ {\hat n}_{nl} \mp \hat{\delta}_{ln}\right\},\quad
\label{eq:u_pm-def}
\end{eqnarray}
The final step is to perform an ensemble average of Eq.~(\ref{eq:eom-nij_final}), where we denote the 
average of an operator by $\langle {\hat A}\rangle \equiv A$ and express the two-operator average via the fluctuations
 $\langle {\hat A}{\hat B} \rangle = A B + \langle \delta{\hat A}\delta{\hat B} \rangle$, where $\delta \hat{A} = {\hat A} - A$.
With these definitions, the ensemble average of Eq.~(\ref{eq:eom-nij_final}) is given by
\begin{eqnarray}
i\partial_t {\bf n}(t) - 
\left[ {\bf n}(t),{\bf t}^* + {\bf v}^*_{\rm H}(t) + {\bf U}^{\pm}_{\rm H}(t)\right]_- =
%\\
%&=& 
\left\langle \left[ \delta{{\bf \hat n}}(t),\delta{\hat {\bf U}}^{\pm}_{\rm H}(t)\right]_- \right\rangle
\equiv {\bf I}^{\rm cor}(t).
\label{eq:average_eom_n_1} 
\end{eqnarray}
The commutator on the left describes the mean-field dynamics (denoted $I^{\rm MF}$ in the main text), the term on the right is related to all contributions beyond mean field, i.e. to correlations.

Equation (\ref{eq:average_eom_n_1}) corresponds to Eq. (4) in the main text, where the short operator notation $U_{\rm HF}$ is being used for the Heisenberg matrix ${\bf U}^{\pm}_{\rm H}$, and $\rho$ is being used for the Heisenberg matrix ${\bf n}$. 
While Eq.~(\ref{eq:average_eom_n_1}) is an equation for ensemble averaged quantities (corresponding to the left part of Fig. 1 of the main text), Eq. (4) of the main text replaces the ensemble average by a ``classical'' average $\overline{n}$ over random realizations $n^{(n)}$, as is illustrated by the right part of Fig. 1 and further discussed in the main text.

%References: